\documentclass{PoS}
\title{3D hydrodynamical {\tt CO$^5$BOLD} model atmospheres of late-type giants: chemical abundances from molecular lines}

\ShortTitle{3D hydrodynamical {\tt CO$^5$BOLD} model atmospheres of late-type giants}

\author{\speaker{A. Ivanauskas}\\
        Vilnius University Astronomical Observatory, M.K. \v Ciurlionio 29, Vilnius LT-10222, Lithuania\\
        Vilnius University Institute of Theoretical Physics and Astronomy, A. Go\v stauto 12, Vilnius LT-01108, Lithuania\\
        E-mail: \email{augustinas.ivanauskas@tfai.vu.lt}}

\author{A. Ku\v cinskas\\
        Vilnius University Institute of Theoretical Physics and Astronomy, A. Go\v stauto 12, Vilnius LT-01108, Lithuania\\
        Vilnius University Astronomical Observatory, M.K. \v Ciurlionio 29, Vilnius LT-10222, Lithuania\\
        E-mail: \email{arunaskc@itpa.lt}}

\author{H.-G. Ludwig\\
        ZAH Landessternwarte Konigstuhl, D-69117 Heidelberg, Germany\\
        E-mail: \email{hludwig@lsw.uni-heidelberg.de}}

\author{E. Caffau\\
        ZAH Landessternwarte Konigstuhl, D-69117 Heidelberg, Germany\\
        E-mail: \email{ecaffau@lsw.uni-heidelberg.de}}

\abstract{We investigate the influence of convection on the formation of
  molecular spectral lines in the atmospheres of late-type giants. For this
  purpose we use the 3D hydrodynamical {\tt CO$^5$BOLD} and classical 1D {\tt
    LHD} stellar atmosphere codes and synthesize a number of fictitious lines
  belonging to a number of astrophysically relevant molecules, C$_2$, CH, CN, NH, OH. We find that differences between the abundances obtained from
  molecular lines using the 3D and 1D model atmospheres are generally small at
  [M/H]=0.0, but they quickly increase at sub-solar metallicities and may reach $\sim-0.9$\,dex at [M/H]=--3.0. The 3D--1D abundance corrections show a significant dependence on the spectral line parameters, such as wavelength and excitation potential. Our comparison, therefore, points to a complex interplay between the spectral line formation and convection that should be properly adressed in stellar abundance analysis.}

\FullConference{11th Symposium on Nuclei in the Cosmos, NIC XI\\
		July 19-23, 2010\\
		Heidelberg, Germany}

\begin{document}
\section{Introduction}
Late-type giant stars are important tracers of intermediate and old (>1 Gyr) stellar populations. They
are luminous and can be detected in distant stellar systems of the Milky Way and Local Group galaxies
and may thus provide important clues about the origins of different chemical species in the Universe.
The latter can be achieved using accurate chemical abundances determined with realistic stellar model
atmospheres and spectral synthesis calculations. However, suitability of the currently used 1D model
atmospheres for achieving this goal has been seriously questioned with the advent of the new generation
3D hydrodynamical model atmospheres. Doubts are raised by the fact that convection
influences strongly the average temperature stratification and produces horizontal temperature and density fluctuations that are absent in the 1D models, which may thus influence the physical conditions in the line forming regions (e.g., [3]).

\section{Aim and tools}
In this work we investigate the influence of convection on the formation properties of
molecular lines in the atmospheres of late-type giants. We used 3D and 1D stellar model atmospheres
to perform the analysis of artificial spectral lines corresponding to a number of astrophysically
important molecules. The goal of this study was to derive quantitative constraints on the abundance
estimates obtained from molecular lines using the 3D and 1D stellar model atmospheres. We will further call the difference between the abundances obtained using the two kinds of atmosphere models a 3D--1D abundance correction (e.g., [4], [2]).

Our analysis was based on the 3D and 1D model atmospheres calculated using the
{\tt CO$^5$BOLD} and {\tt LHD} stellar atmosphere codes, respectively. Both
stellar atmosphere codes share the same micro-physics (opacities, equation of
state) and atmospheric parameters (Table~\ref{tab1}). Using the 3D and 1D
model atmospheres we synthesized a set of fictitious spectral lines
corresponding to several molecules: C$_2$, CH, CN, NH,
OH. Spectral line synthesis was performed with the Linfor3D code under the assumption
of LTE (for the description of all three codes see, e.g., [1], [5]). The wavelengths and excitation potentials of the fictitious
spectral lines were 4000, 8500, 16000 {\AA} and 0, 2 eV,
respectively. The bluest wavelength was chosen as representative of molecular
lines observed in this spectral region (e.g., CH G band at $\sim4300${\AA}). The 8500\,{\AA} and 16000\,{\AA}
wavelengths coincide with the maximum and minimum of the H$^-$ ion bound-free
opacity, respectively.

The 3D--1D abundance corrections obtained for a given spectral line with the
3D and 1D model atmospheres were derived using the curves of growth. Only weak
spectral lines ($\sim$ <5 m{\AA}) were used in the analysis in order to
eliminate the influence of the microturbulent velocity in the 1D models.

\begin{table}
\vspace*{-.1 cm}
\begin{tabular}{lccccc}
\hline

Model & Size($x$,$y$,$z$) [Mm$^3$] & Grid points ($n_{\rm x}$, $n_{\rm y}$, $n_{\rm z}$) & $T_{\rm eff}$ [K] & $\log{g}$ [cgs] & [M/H]\\
\hline
d3t50g25mm00  & 573 x 573 x 243 & 160 x 160 x 200  & 4970 & 2.5 &  0.0 \\
d3t50g25mm10  & 573 x 573 x 245 & 160 x 160 x 200  & 4990 & 2.5 & -1.0 \\
d3t50g25mm20  & 584 x 584 x 245 & 160 x 160 x 200  & 5020 & 2.5 & -2.0 \\
d3t50g25mm30  & 573 x 573 x 245 & 160 x 160 x 200  & 5020 & 2.5 & -3.0 \\
\hline
\end{tabular}
\caption{Parameters of the 3D {\tt CO$^5$BOLD} models of late-type giants used in the present study.}
\label{tab1}
\end{table}

\section{Results and discussion}
\subsection{3D--1D abundance corrections}
\begin{figure}
\begin{center}
\resizebox{0.86\textwidth}{!}{\includegraphics{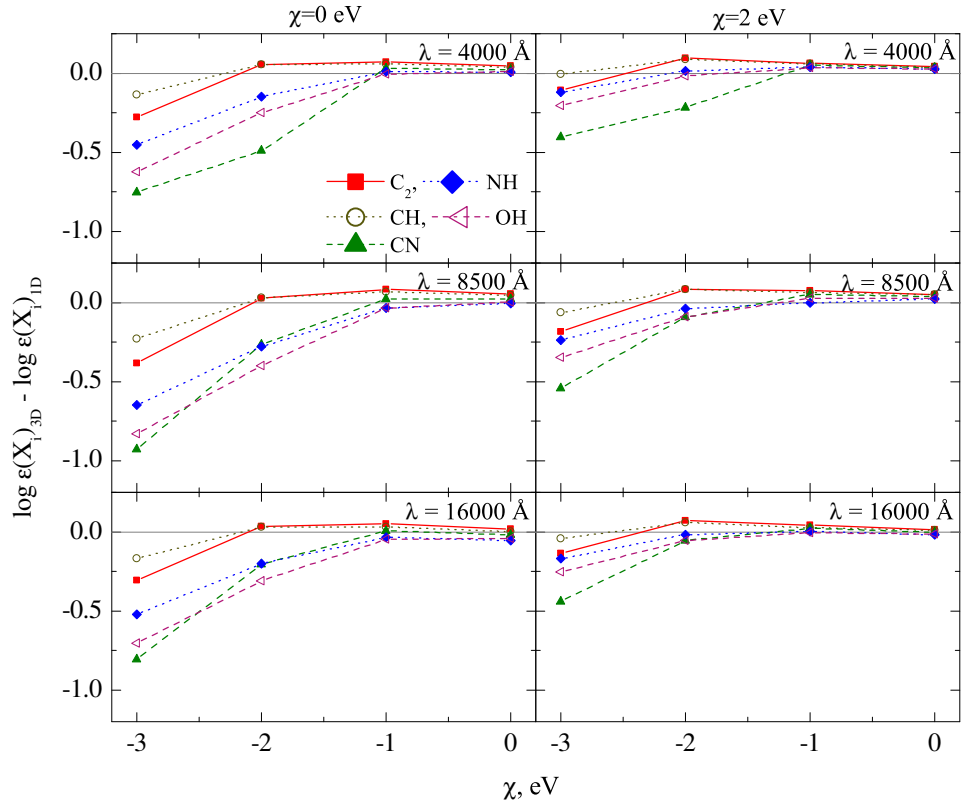}}
\caption{The 3D--1D abundance corrections for different molecular species derived using fictitious lines with excitation
potentials $\chi$ = 0, 2 eV, plotted versus metallicity at three different wavelengths, 4000, 8500, 16000 {\AA}.}
\label{fig1}
\end{center}
\vspace*{-.1cm}
\end{figure}

The 3D--1D abundance corrections derived at four different metallicities from various molecular lines with different wavelength, $\lambda$, excitation potential of the lower level $\chi$ are shown in Fig.~\ref{fig1}. The influence of convection is generally largest at lowest metallicity ([M/H] = --3.0). At a given metallicity, the size of 3D--1D abundance corrections decreases with increasing excitation potential of a given spectral line, $\chi$ (Fig.~\ref{fig1}). The 3D--1D abundance corrections at low metallicities differ significantly in their magnitude which ranges from $\sim$ --0.1\,dex in case of CH to $\sim$ --0.9\,dex in case of CN.


\begin{figure}
\vspace*{-.1 cm}
\begin{center}
\resizebox{1.00\textwidth}{!}{\includegraphics{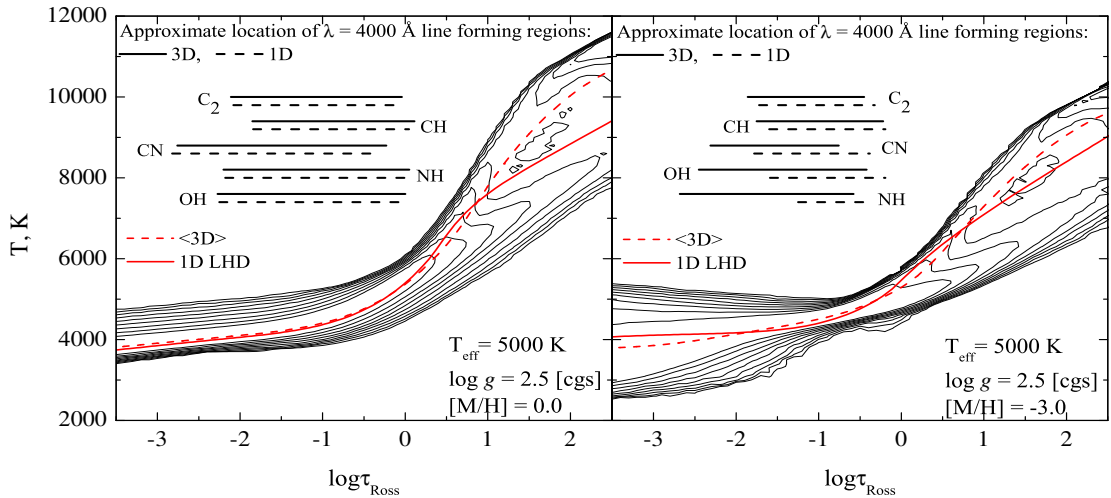}}
\caption{Temperature profiles of the 3D hydrodynamical and classical 1D models of the late type giant plotted versus Rosseland optical depth at two different metallicities, [M/H] = 0.0 (left) and [M/H] = --3.0 (right). The contour plot shows the density of the individual temperature profiles of the 3D model. The dashed line represents the temperature profile of the temporally and spatially averaged $\langle{\rm 3D}\rangle$ model, while the solid line represents the corresponding 1D model. Horizontal black lines in the upper part of the plots show the approximate location where various molecular lines form in the 3D and 1D models (solid and dashed lines, respectively).}
\label{fig2}
\end{center}
\vspace*{-.3cm}
\end{figure}

\subsection{Thermal structures in the 3D models}

The comparison of the temperature profiles of the 3D model with the corresponding temperature stratifications of the 1D {\tt LHD} model and the temporarily and spatially averaged 3D hydrodynamical (hereafter, $\langle{\rm 3D}\rangle$) model is shown in Fig.~\ref{fig2}. At solar metallicity the 1D {\tt LHD} model closely traces the most probable temperature profile of the 3D model. However, at [M/H] = --3.0 and beyond $\log \tau_{\rm Ross}\sim-3.5$ the 3D model is cooler than the corresponding 1D model (Fig.~\ref{fig2}).


The formation of different molecular lines occurs at different optical depths, both at solar and sub-solar metallicities (Fig.~\ref{fig2}). However, since the differences between the temperature stratifications of the averaged 3D and 1D models are small at [M/H]=0.0 the corresponding 3D--1D abundance corrections are minor. The situation changes at lower metallicities where even small differences in the location of molecular line formation are associated with significant changes in the local temperatures, due to rapidly increasing differences between the temperature profiles of the averaged 3D and 1D models in the outer atmosphere. This may explain why, for example, the 3D--1D abundance corrections are smallest in the case of C$_2$ and CH, which form deepest in the 3D atmosphere and are least affected by differences in the temperature profiles.

\subsection{Comparison with the work of other authors}

\begin{figure}
\vspace*{-.1 cm}
\begin{center}
\resizebox{1.0\textwidth}{!}{\includegraphics{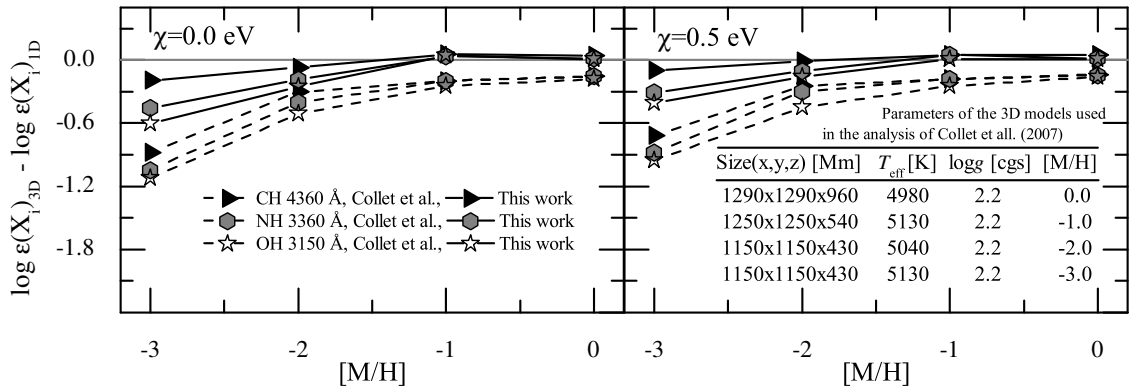}}
\caption{The 3D--1D abundance corrections for CH 4360 Å, NH 3360 Å, OH 3150 Å lines. Dashed lines are results of [3], solid lines represent calculations obtained in this study. Table inside the figure gives the parameters of the 3D models used by [3].}
\label{fig3}
\end{center}
\vspace*{-.01cm}
\end{figure}

The only similar study where the effects of convection on spectral line
formation in late-type giant stars were analyzed was done
by [5]. The results obtained in the two studies agree in a sense
that largest 3D--1D abundance corrections are seen in case of the most metal-
poor atmosphere models (Fig.~\ref{fig3}). However, the magnitude of the
abundance corrections obtained in our study is significantly smaller. The
origin of these differences is not fully clear yet. In part, it
might be related to the fact that the $\langle{\rm 3D}\rangle$--1D temperature differences are larger in [5], particularly at lowest metallicites, but one should keep in mind that gravities in the two sets of models are also slightly different.

\section{Conclusions}
Convection alters significantly the thermodynamical structure of late-type
giant atmospheres, especially in the outer layers, and its influence grows
larger with decreasing metallicity. At lowest metallicities, the temperature in
the outer layers of the averaged 3D model is significantly lower than in the
corresponding 1D model atmosphere. This leads to significant differences
between the molecular abundances derived using the 3D and 1D model
atmospheres: for example, in case of CN they may be as large as --0.9 dex at [M/H] = --3.0. The 3D--1D corrections predicted for the individual molecules are also markedly different, with
the largest spread at lowest metallicities.  The strong dependence of the
3D--1D abundance correnctions on oscillator strength, wavelength, and excitation
potential of particular spectral line clearly exposes the limitations of the
1D models in stellar abundance work, especially when studying metal-poor stellar populations.

\end{document}